# Realistic simulation of reflection high-energy electron diffraction patterns for two-dimensional lattices using Ewald construction

Chong Liu,[1,2,a)] Kai Chang,[1] and Ke Zou[2,3]

[1]Beijing Academy of Quantum Information Sciences, Beijing 100193, China
[2]Quantum Matter Institute, University of British Columbia, Vancouver, British Columbia V6T 1Z4, Canada
[3]Department of Physics & Astronomy, University of British Columbia, Vancouver, British Columbia V6T 1Z1, Canada

[a)]Electronic mail: liuchong@baqis.ac.cn

Reflection high-energy electron diffraction (RHEED) is a powerful tool for characterizing crystal surface structures. However, the setup geometry leads to distorted and complicated patterns, which are not straightforward to link to the real-space structures. A program with a graphical user interface is provided here to simulate the RHEED patterns. Following the Ewald construction in the kinematic theory, we find out the exact geometric transformation in this model that determines the positions of diffraction spots. The program can deal with many forms of surface structures, including surface reconstructions or domains. The simulations exhibit great agreement with the experimental results in various cases. This program will benefit the structure analysis in thin film growth and surface science studies.

# I. INTRODUCTION

Reflection high-energy electron diffraction (RHEED) is a widely used technique in vacuum systems for thin-film growth such as molecular beam epitaxy (MBE) and pulsed laser deposition (PLD). The incident electron beam with the energy of 10–100 keV comes to the crystal surface at a grazing angle and the forward scattering electrons are detected on a screen. It provides information about the surface crystal structure, flatness, domains, phase separation, film growth mode, layer period and so on.[1] For the single scattering process, the RHEED pattern can be viewed as a projection of the reciprocal lattice, yet heavily distorted due to the grazing geometry. As a result, it is difficult to extract the lattice structure immediately from the experimental data.[2] In practice, one usually needs simulation of candidate structures to predict or compare with the actual RHEED results.

Simulating the RHEED pattern is relatively complicated and could be CPU costly for computers. As a compromise, for instance, a previous approach had to simplify the Ewald sphere as a plane.[3] Consequently, the simulation only captured the diffractions in the zeroth Laue zone and displayed as long streaks in contrast to spotted patterns as they should be. It is necessary to preserve the Ewald sphere in the model, essentially taking the complete reciprocal array into account, to properly reflect the detailed lattice structure and the more complicated reconstructions and domains. In this article, we provide an open-source program that utilizes the exact Ewald construction and accomplishes realistic RHEED simulation for two-dimensional (2D) crystal surfaces. Written in Wolfram *Mathematica*, the program has a graphical user interface (GUI) and is time-efficient on personal computers (see supplementary material at [URL will be inserted by AIP Publishing] for the program, source code and Manual). The core theoretical model and potential applications will be displayed.

## II. EXPERIMENT

The experimental data were acquired with a RHEED gun from STAIB Instruments, installed on a Veeco GenXplor MBE system. The beam energy was 10 keV, the emission current was 0.2 µA and the grazing angle was around 2.5°. The MgO film was grown on a MgO(001) substrate at 600°C by depositing Mg atoms at the oxygen pressure of $1.2 \times 10^{-6}$ Torr. The Ge(111) substrate was etched in 20% HCl solution for 20 minutes and then loaded in the MBE chamber. After annealing at 490°C in vacuum for 30 minutes, Ge atoms were deposited on the substrate until clear RHEED patterns were obtained. $SrTiO_3$(001) substrate was etched in boiled water for 1 hour and 10% HCl for 45 minutes, and then annealed in a tube furnace in flowing oxygen gas at 1120°C for 4 hours. After loading in the MBE chamber, it was degassed at 620°C for 30 minutes to obtain the clean surface.

## III. THEORETICAL MODEL

### A. Ewald construction

The kinematic theory of the RHEED has been well established.[4,5] Considering a 2D lattice in the *x-y* plane whose primitive cell is described by basis vectors **a** and **b**, the corresponding reciprocal vectors are

$$\mathbf{a}^* = 2\pi \frac{\mathbf{b} \times \hat{\mathbf{z}}}{|\mathbf{a} \times \mathbf{b}|} \quad (1)$$

$$\mathbf{b}^* = 2\pi \frac{\mathbf{a} \times \hat{\mathbf{z}}}{|\mathbf{a} \times \mathbf{b}|} \quad (2)$$

where $\hat{\mathbf{z}}$ is the unit vector normal to the lattice plane. A reciprocal vector is obtained as

$$\mathbf{G}(hk) = h\mathbf{a}^* + k\mathbf{b}^*. \qquad (3)$$

$h$ and $k$ are Miller indices.

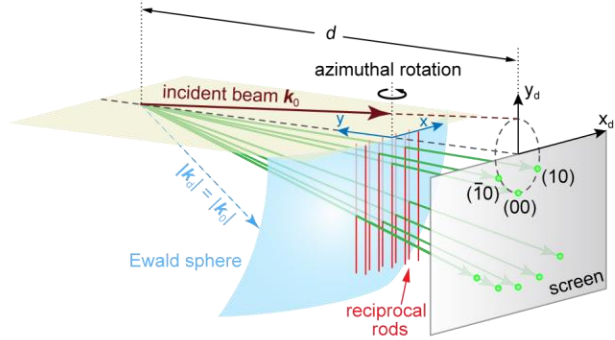

FIG. 1. Schematic geometry of RHEED with the Ewald construction. The graph integrates the real and reciprocal spaces, which are linked by the angle of the wave vectors.

For elastic scattering, energy conservation requires $|\mathbf{k}_d| = |\mathbf{k}_0|$, where $\mathbf{k}_0$ and $\mathbf{k}_d$ are the wave vectors of the incident and diffracted beam, respectively. Hence, the end of any possible $\mathbf{k}_d$ falls on a sphere in the reciprocal space with the radius equal to $|\mathbf{k}_0|$, which is called Ewald sphere, as shown in Figure 1. Furthermore, based on the Laue condition, the diffraction must satisfy

$$\Delta \mathbf{k}^{\parallel} = \mathbf{k}_d^{\parallel} - \mathbf{k}_0^{\parallel} = \mathbf{G}. \qquad (4)$$

The 2D reciprocal lattice extends along the $z$ direction continuously, known as reciprocal rods. If we place the tip of $\mathbf{k}_0$ on one of the rods, according to Eq. (4), $\mathbf{k}_d$ should also lie on the rods. Then the intersection of the Ewald sphere and reciprocal rods determine which diffraction angles are allowed. The diffracted beams in those angles shoot on a screen that is perpendicular to the lattice plane, forming the RHEED patterns.

With the Ewald construction as illustrated in Fig.1, the diffraction process becomes a geometrical problem: a two-step projection of the reciprocal array from the $x$-$y$ plane to the Ewald sphere and then to the $x_d$-$y_d$ plane on the screen.[6] Accordingly, we have derived the coordinate transformation:

$$x = \frac{k_0 x_d}{\sqrt{d^2+x_d^2+y_d^2}} \quad (5)$$

$$y = k_0\left(-\frac{d}{\sqrt{d^2+x_d^2+y_d^2}} + \cos\theta\right), \quad (6)$$

where $d$ is the distance between the sample and the screen and $\theta$ is the grazing angle of the incident beam. This transformation is the key operation in the model and the code, which guarantees realistic simulations of RHEED patterns.

## B. STRUCTURE FACTOR

The Ewald construction determines the position of the diffracted spots while the structure factor determines the intensity of each diffraction. In most real crystals, there are multiple atoms in the unit cell, and the interference between them will modulate the intensity of the diffracted beam. The structure factor is obtained as

$$F(\boldsymbol{G}) = \sum_m f_m \cdot e^{i\boldsymbol{G}\cdot\boldsymbol{r}_m}, \quad (7)$$

where $f_m$ is the electron scattering factor for the atom,[7] $\boldsymbol{r}_m$ is the position of the atom and the sum is over all atoms in the unit cell. The diffraction intensity is $I(\boldsymbol{G}) = |F(\boldsymbol{G})|^2$.

In experiments, the reciprocal rods are broadened due to finite coherence length and finite domain size. We use 2D Gaussian function with the intensity proportional to $I(\boldsymbol{G})$ to model the intensity distribution of every reciprocal rod.

## IV. PROGRAM STURCTURE

The program runs in the following workflow:

(1) Create single or multiple lattices based on the user's settings. Assign two basis vectors to each lattice.

(2) From user-input coordinates of the basis vectors, calculate the reciprocal basis vectors and generate a reciprocal array $\boldsymbol{G}(hk)$ [Eqs. (1)-(3)]. If an azimuthal rotation is required, then a rotational transformation is applied to all vectors.

(3) From user-input $\boldsymbol{r}_m$ and $f_m$, calculate the diffraction intensity $I(hk) = |F(hk)|^2$ based on Eq. (7).

(4) Create a 2D intensity distribution function $L(x,y)$ in the *x-y* plane by summing up 2D Gaussian functions centered at every $\boldsymbol{G}(hk)$ with intensity $I(hk)$. If multiple superposed lattices are defined in the beginning, their $L(x,y)$ maps are calculated independently and summed up. The map of $L(x,y)$ is approximately a simulation of low energy electron diffraction (LEED) pattern.

(5) Apply broadening along *x* and *y* directions of the Gaussian functions at all $\boldsymbol{G}(hk)$. Transform the intensity map $L(x,y)$ to the $(x_\mathrm{d}, y_\mathrm{d})$ coordinates by Eqs. (5)-(6). The resulting map is plotted as the simulation of the RHEED pattern.

## V. APPLICATIONS AND DISCUSSION

The program contains a GUI for fully interacting operations. Besides the lattice structures, many experimental parameters are free to customize, such as beam energy, incident angle, sample azimuth, spot broadening and overall intensity. Furthermore, it supports multiple sets of lattices in a single simulation, so that the simulation can cover more complicated cases in practice, including reconstruction, phase separation and rotational domains. For instruction on the use of the program, refer to the *Manual* in the supplementary material at [URL will be inserted by AIP Publishing].

Here we display various examples to demonstrate the capability and flexibility of the program. The input parameters for all examples are displayed in Table S1 in the supplementary material at [URL will be inserted by AIP Publishing].

For some crystals, the structure factors could vanish for certain Miller indices, leading to zero intensity. MgO has a face-centered cubic (fcc) structure. The condition to have non-zero diffraction intensity is that the Miller indices are all odd or all even. As shown in Figs. 2(a, b), the (00), (20) and (11) diffractions are visible while (10) is missing. Our simulations give the same results [Figs. 2(c, d)] when considering the structure factors.

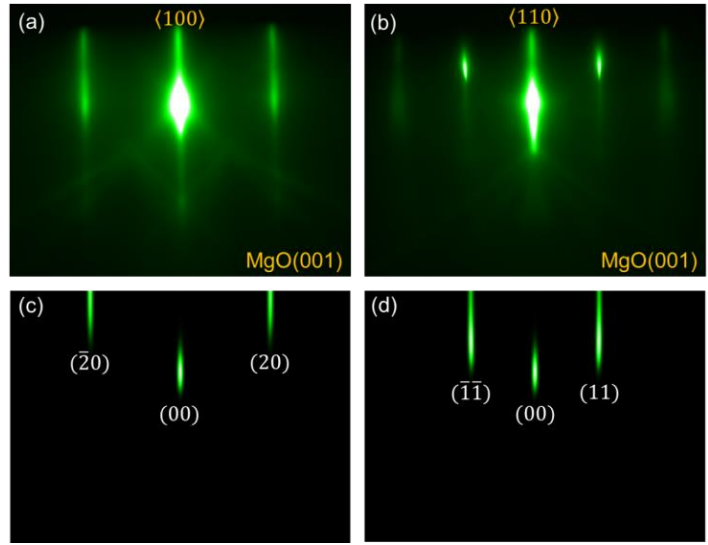

FIG. 2. (a, b) Experimental RHEED images of MgO(001) surface with the incident beam along ⟨100⟩ and ⟨110⟩ directions, respectively. (c, d) Simulated patterns corresponding to (a, b), respectively, which reproduce the cancellation of the (10) diffraction due to the fcc structure.

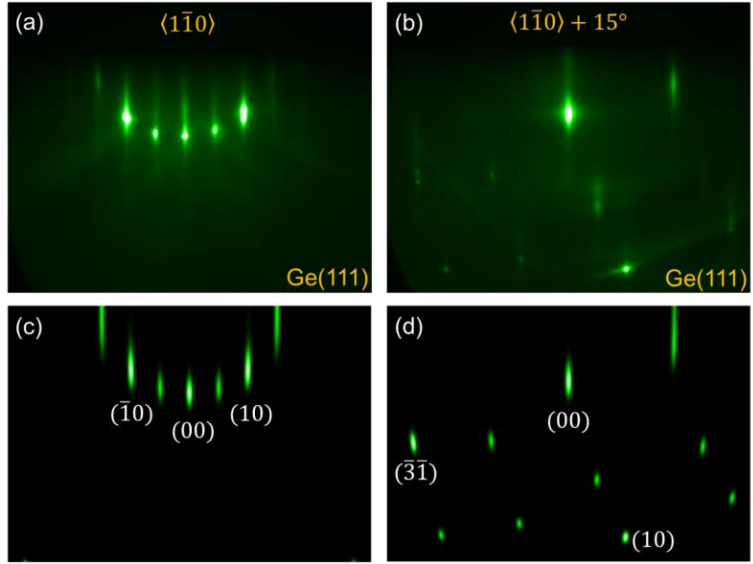

FIG. 3. (a, b) Experimental RHEED images of 2×2 reconstructed Ge(111) surface with the incident beam along ⟨1$\bar{1}$0⟩ direction and 15 ° azimuth from ⟨1$\bar{1}$0⟩ direction, respectively. (c, d) Simulated patterns corresponding to (a, b), respectively.

An atomically flat surface often has reconstructions. Fig. 3(a) is the RHEED pattern of Ge(111) surface. One can tell from the half-order diffraction spots that the surface is reconstructed, but it's insufficient to pin down the exact reconstruction type since the image only contains one row of the reciprocal lattice.[8] By rotating the sample azimuth, the reciprocal rods are scanned over by the Ewald sphere (Fig. 1), thus the full structure information could be reflected in RHEED. A correct model should reproduce the diffraction exclusively for arbitrary azimuth angles such as that in Fig. 3(b). We create a hexagonal lattice overlaid with a 2×2 lattice in our program. The simulation results [Figs. 3(c, d)] agree with the experimental ones very well. To verify the uniqueness of the explanation, we also create and simulate 2×1 and 2×3 reconstructions for these angels (see Fig. S1 in supplementary material at [URL will be inserted by AIP Publishing]). Although all 2×$n$ reconstructions possess similar half-order diffractions in the 0$^{th}$ Laue zone in ⟨1$\bar{1}$0⟩ direction, they are distinguished by the number of spots in the intermediate Laue zones, which are better manifested in the asymmetric angle at $\langle 1\bar{1}0\rangle + 15°$. Hence, the experimental phase is unambiguously proved to be 2×2 reconstruction, which usually occupies small partial areas on the Ge(111) surface [9] unless it is stabilized by highly dense steps.[10,11] This testing shows that the program works for any incident directions and hence is useful for recognizing uncommon surface structures.

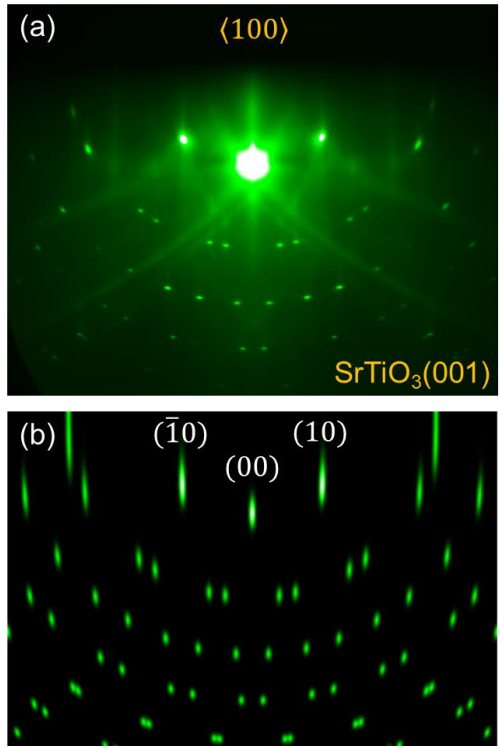

FIG. 4. (a) Experimental RHEED image of SrTiO$_3$(001) surface with $(\sqrt{13} \times \sqrt{13})$R33.7° reconstructions. The incident beam is along the $\langle 100\rangle$ direction. (b) Simulated RHEED pattern corresponding to (a), which considers two domains (rotating $\pm 33.7°$) of the $\sqrt{13} \times \sqrt{13}$ reconstruction. The grazing angle is set to be 2.6°. The simulation reproduces all diffraction spots seen in the experiment.

The multi-lattice functionality makes it suitable for modelling domains. The $\sqrt{13} \times \sqrt{13}$ reconstruction of $SrTiO_3$(001) has a large supercell that is not aligned with the original 1×1 lattice. The supercell rotates 33.7° or −33.7° with respect to the 1×1 lattice with equal probability, forming domains that induce complex patterns in the RHEED [Fig. 4(a)].[12] Based on the aforementioned two rotated lattices, our program yields a decent simulation for the reconstruction, which gives a 2D map including the full spot arrays in the intermediate Laue zones [Fig. 4(b)]. Every single spot is reproduced at the exact position, exhibiting high precision of the simulation.

With such good performance of the program, it will be able to determine unknown reconstruction types in a definite way. Despite the complication in calculation and visualization, the evaluation of this simulation only costs a few seconds on a personal computer (see supplementary material at [URL will be inserted by AIP Publishing] for information of testing system and running time). A similar example yielded by the same algorithm has been involved in previous research work.[13] In the same way, it also works for domains consisting of multiple phases or rotations.[14]

This program only considers 2D layer meshes, which represent well for most flat surfaces, as has been demonstrated above, since the number of penetrating elastic electrons decays exponentially with increasing depth. Still, this simplification will miss some fine features that originate from multiple-layer diffraction, finite coherent area, islands, steps, defects and disorders. Three-dimensional models and interlayer interference might be included in future updates. More features, although more challenging, remain to be modeled, such as Kikuchi lines that are caused by multiple scatterings of the electrons.

Applying machine learning on RHEED data is a rising trend in surface and thin-film sciences.[15-18] One potential application of this program is to generate predictive RHEED images for machine learning codes. For instance, a large number of various artificial RHEED images can be generated to feed into training for a neural network, which can then be used for data identification or classification.

## VI. CONCLUSIONS

In summary, we developed a program to realize the simulation of RHEED patterns of 2D crystal surfaces. Based strictly on the Ewald construction, the algorithm works for single crystal, reconstructions and multiple domains. Most experimental parameters are taken into account so that the best agreement can be achieved. With both high performance and efficiency, the program will enhance the power of RHEED technique and have wide application in research.

## ACKNOWLEDGMENTS

The authors thank Bruce Davidson at UBC for the discussions on the principles of RHEED. This work was undertaken, thanks in part to funding from the Max Planck-UBC-UTokyo Centre for Quantum Materials and the Canada First Research Excellence Fund, Quantum Materials and Future Technologies Program. The work at UBC was also supported by the Natural Sciences and Engineering Research Council of Canada (NSERC) and the Canada Foundation for Innovation (CFI). K. C. acknowledges the funding from National Natural Science Foundation of China under Grant No. 12074038 and No. 92165104.

# AUTHOR DECLARATIONS

**Conflict of Interest**

The authors have no conflicts to disclose.

# DATA AVAILABILITY

The data that supports the findings of this study are available within the article and its supplementary material.

Supplementary Material for

# Realistic simulation of reflection high-energy electron diffraction patterns for two-dimensional lattices using Ewald construction

Chong Liu,[1,2,a)] Kai Chang,[1] and Ke Zou[2,3]

[1]Beijing Academy of Quantum Information Sciences, Beijing 100193, China

[2]Quantum Matter Institute, University of British Columbia, Vancouver, British Columbia V6T 1Z4, Canada

[3]Department of Physics & Astronomy, University of British Columbia, Vancouver, British Columbia V6T 1Z1, Canada

[a)]Electronic mail: liuchong@baqis.ac.cn

TABLE S1. Input parameters for the simulations displayed in main text.

|  | **MgO(001)** | **Ge(111)-(2×2)** | **STO(001)-($\sqrt{13} \times \sqrt{13}$)** |
|---|---|---|---|
| Number of lattices | 1 | 2 | 2 |
| Lattice base vectors: ($x_1,y_1$), ($x_2,y_2$) | (1,0), (0,1) | ($\sqrt{3}/2$,-1/2), (0,1) | (2,3), (3,-2) |
|  |  | ($\sqrt{3}$,-1), (0,2) | (3,2), (2,-3) |
| Structure factors: ($r_a,r_b,f$) | (0,0,1) | (0,0,1) | (0,0,1) |
|  | (0.5,0.5,1) |  |  |
| Lattice unit (Å) | 4.2 | 4 | 3.9 |
| Grazing angle (°) | 3 | 3 | 2.6 |
| Rotation (°) | <100>: 0<br><110>: 45 | <1$\bar{1}$0>: 0<br><1$\bar{1}$0>+15°: 15 | <100>: 0 |
| Beam $E$ (keV) | 10 | 10 | 10 |
| Broaden x, y | 1, 1.5 | 1, 1 | 0.8, 0.7 |

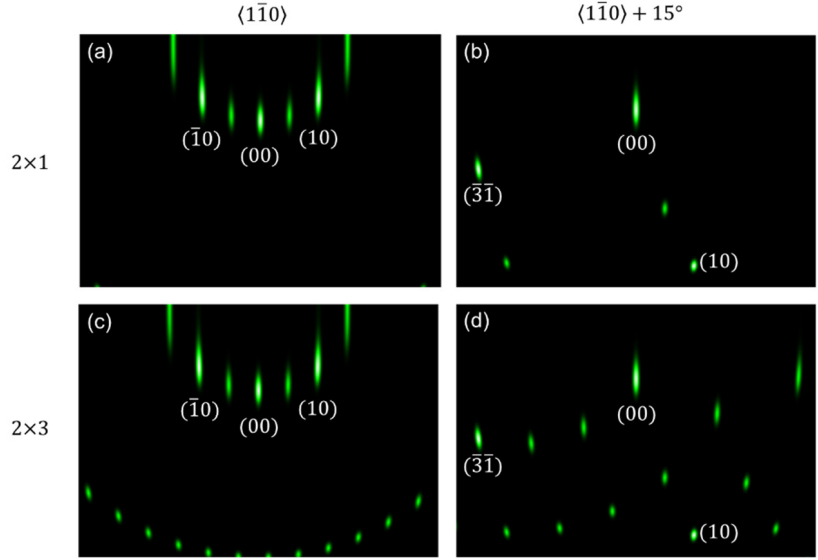

FIG. S1. (a, b) Simulated patterns of 2×1 reconstructed Ge(111) surface with the incident beam along $\langle 1\bar{1}0 \rangle$ direction and 15° azimuth from $\langle 1\bar{1}0 \rangle$ direction, respectively. (c, d) Simulated patterns of 2×3 reconstructed Ge(111) surface with the incident beam along $\langle 1\bar{1}0 \rangle$ direction and 15° azimuth from $\langle 1\bar{1}0 \rangle$ direction, respectively.

## Running time tests

The code running time increases with increasing number of lattices and reciprocal spots in the plotting range. Taking the most complicated case in this article, $\sqrt{13} \times \sqrt{13}$ reconstruction of STO(001), as an example, the computer system information and testing results are as following.

CPU: Intel(R) Core™ i5-10210U CPU @ 1.60GHz

RAM: 16 GB, 2133 MHz

Operation system: Windows 11

Open up main program interface: 1.0 second

Update reciprocal lattice plot: 9.1 seconds (For preview only. Unnecessary for RHEED simulation)

Generate RHEED simulation and display in a new window: 7.7 seconds

Update the simulation after changing parameters: 2.9 seconds